\documentclass[fleqn,10pt]{wlpeerj}
\usepackage{graphicx}
\usepackage{url}
\usepackage{tabularx}
\usepackage{booktabs}
\usepackage{longtable}
\usepackage{xcolor}
\usepackage{mdframed}
\usepackage[most]{tcolorbox}
\usepackage{adjustbox}

\newcommand{\labeledfbox}[2]{\noindent\fbox{\begin{minipage}{\textwidth} 
    \textbf{#1}\\ #2
  \end{minipage}}\label{#1}}

\title{Harnessing multiple LLMs for Information Retrieval: A case study on Deep Learning methodologies in Biodiversity publications}

\author[1,2,3]{Vamsi Krishna Kommineni}
\author[1,2,4]{Birgitta König-Ries}
\author[5]{Sheeba Samuel}
\affil[1]{Heinz Nixdorf Chair for Distributed Information Systems, Friedrich Schiller University Jena, Jena, Germany}
\affil[2]{German Centre for Integrative Biodiversity Research (iDiv) Halle-Jena-Leipzig, Leipzig, Germany}
\affil[3]{Max Planck Institute for Biogeochemistry, Jena, Germany}
\affil[4]{Michael Stifel Center Jena, Jena,  Germany}
\affil[5]{Distributed and Self-organizing Systems, Chemnitz University of Technology, Chemnitz, Germany}
\corrauthor[1]{Vamsi Krishna Kommineni}{vamsi.krishna.kommineni@uni-jena.de}

\begin{abstract}
Deep Learning (DL) techniques are increasingly applied in scientific studies across various domains to address complex research questions. 
However, the methodological details of these DL models are often hidden in the unstructured text. As a result, critical information about how these models are designed, trained, and evaluated is challenging to access and comprehend.
To address this issue, in this work, we use five different open-source Large Language Models (LLMs): Llama-3 70B, Llama-3.1 70B, Mixtral-8x22B-Instruct-v0.1, Mixtral 8x7B, and Gemma 2 9B in combination with Retrieval-Augmented Generation (RAG) approach to extract and process DL methodological details from scientific publications automatically. We built a voting classifier from the outputs of five LLMs to accurately report DL methodological information. 
We tested our approach using biodiversity publications, building upon our previous research. To validate our pipeline, we employed two datasets of DL-related biodiversity publications:
a curated set of 100 publications from our prior work and an additional set of 364 publications from the Ecological Informatics journal. 
Our results demonstrate that the multi-LLM, RAG-assisted pipeline enhances the retrieval of DL methodological information, achieving an accuracy of 69.5\% (417 out of 600 comparisons) based solely on textual content from publications. This performance was assessed against human annotators who had access to code, figures, tables, and other supplementary information.  
Although demonstrated in biodiversity, our methodology is not limited to this field; it can be applied across other scientific domains where detailed methodological reporting is essential for advancing knowledge and ensuring reproducibility. This study presents a scalable and reliable approach for automating information extraction, facilitating better reproducibility and knowledge transfer across studies.
\end{abstract}

\begin{document}

\flushbottom
\maketitle
\thispagestyle{empty}

\section*{Introduction}
Deep Learning (DL) has become a cornerstone in numerous fields, revolutionizing how complex data is analyzed and interpreted.
From healthcare and finance to autonomous systems and natural language processing, DL techniques have delivered groundbreaking results.
However, as the adoption of DL continues to grow, there is an increasing recognition of a critical shortcoming: the limited availability of detailed methodological information in scientific literature \citep{waide2017demystifying,stark2018before,samuel2020machine, pineau2021improving, gundersen2022machine}. 
This gap presents significant challenges for researchers and practitioners who seek to understand, replicate, and build upon existing studies \citep{feng2019checklist, gpai2022biodiversity}.
Past research has emphasized the need to make primary data and clear metadata available to support transparency \citep{michener1997nongeospatial, whitlock2011data}.

A DL pipeline is a structured process for training and deploying DL models, starting with data collection and preprocessing tasks like cleaning, normalization, and transformation \citep{el2020deep}. After preparing the data, the pipeline moves to model selection, where an appropriate architecture is chosen based on model complexity and problem type. The selected model is then trained on preprocessed data, fine-tuning through specific optimization algorithms and hyperparameter configurations. Once trained, the model’s performance is evaluated on test data to ensure reliable, unbiased results. The final step involves deploying the model for real-world use or further refinement.

For a DL pipeline to be reproducible, detailed documentation at each stage is essential \citep{pineau2021improving}. This includes logging data collection methods, preprocessing steps, model architecture configurations, hyperparameters, and training details, as well as performance metrics and test datasets. Additionally, maintaining records of software libraries, hardware, frameworks, and versions used is critical for the accurate replication of the study.
Without access to such crucial information, stakeholders—including academics, industry professionals, and policymakers—face significant challenges in validating study outcomes or advancing research in meaningful ways. In areas like healthcare, finance, and autonomous systems, where DL applications influence real-world decisions, the absence of methodological transparency can compromise trust in DL models and limit their broader application \citep{haddaway2015poor}. We contend that the same holds true for biodiversity research.

The advent of DL has significantly transformed various domains, including biodiversity research, by enabling advanced methodologies for data analysis and interpretation \citep{AUGUST2020100116}. However, manually extracting relevant deep-learning information from scientific articles remains a labour-intensive and time-consuming process. This challenge affects both the reproducibility of the original studies and the reproducibility of secondary analyses aimed at understanding the methods employed. Traditional manual retrieval methods can be inconsistent, as the perspective of the annotators often varies based on their task interpretation and domain knowledge. These inconsistencies hinder efforts to systematically review or replicate the methodological approaches across studies, highlighting the need for more automated solutions.

To address these challenges, we propose a novel approach that leverages the capabilities of Large Language Models (LLMs) for the automated extraction and processing of DL methodological information from scientific publications. 
LLMs, which are trained on vast amounts of text data, have demonstrated impressive capabilities in natural language understanding and generation. 
Specifically, we employ five open-source LLMs: Llama-3 70B\footnote{\url{https://ai.meta.com/blog/meta-llama-3/}}, Llama-3.1 70B\footnote{\url{https://ai.meta.com/blog/meta-llama-3-1/}}, Mixtral-8x22B-Instruct-v0.1\footnote{\url{https://mistral.ai/news/mixtral-8x22b/}}, Mixtral 8x7B\footnote{\url{https://mistral.ai/news/mixtral-of-experts/}}, and Gemma 2 9B\footnote{\url{https://blog.google/technology/developers/google-gemma-2/}} in combination with Retrieval-Augmented Generation (RAG) approach \citep{lewis2020retrieval}. By utilizing these advanced models, we aim to extract relevant methodological details with greater accuracy and efficiency than manual methods alone.
Our methodology is structured into three critical components: identifying relevant research publications, automatically extracting information through an RAG approach, and converting the extracted textual responses into categorical values for streamlined evaluation.

In this work, we take biodiversity publications as a case study due to the growing popularity of DL methods in biodiversity research and the enormous number of publications using DL for various applications in this domain. Given the importance of biodiversity research and the critical need for transparent sharing of DL information in these studies \citep{gpai2022biodiversity}, we chose this field to demonstrate our approach. However, our methodology is not limited to biodiversity alone; it can be applied to other domains where detailed methodological reporting is essential for advancing scientific knowledge and ensuring reproducibility.

To enhance the reliability of our approach, we developed a voting classifier that aggregates the outputs of these LLMs, ensuring that the reported information is consistent and accurate. This methodology was applied to two distinct datasets of biodiversity publications focused on DL: one consisting of 100 publications from our previous work \citep{ahmed2024evaluating} and another comprising 364 publications from the \textit{Ecological Informatics} journal\footnote{\url{https://www.sciencedirect.com/journal/ecological-informatics}}.

Our approach can help identify gaps in reporting and ensure that critical information about DL methodologies is accessible, thereby enhancing the transparency and reproducibility of research. This paper presents a comprehensive case study on applying multiple LLMs for information retrieval in the context of DL methodologies within biodiversity publications. Through our approach, we aim to contribute to the growing body of research focused on automating information extraction and improving the reproducibility of results in scientific literature. By demonstrating the effectiveness of our pipeline, we hope to pave the way for future research that harnesses advanced AI techniques to further enhance data retrieval and analysis in biodiversity and beyond.
Ensuring reproducibility in LLM applications requires a clear, comprehensive methodology that specifies all critical steps, settings, and model configurations. 
By providing all methodological details transparently, we aim to ensure that our approach can be consistently replicated and applied in future studies, supporting the reliable and reproducible use of LLMs in scientific research.

In the following sections, we provide a detailed description of our study. 
We start with an overview of the state-of-the-art (``Related Work'').
We provide the methodology of our study (``Methods'')
We describe the results of our work (``Results'') and provide a detailed evaluation of our results (``Evaluation'').
We discuss the implications of our study (``Discussion'').
Finally, we summarize the key aspects of our study and provide future directions of our research (``Conclusion'').

\section*{Related Work}
The integration of DL into scientific research has been transformative across a variety of fields, leading to significant advancements in data analysis, pattern recognition, and predictive modelling. 
However, the challenge of adequately documenting DL methodologies has been widely recognized, and several studies have highlighted the importance of transparency and reproducibility in DL research \citep{whitlock2011data, haddaway2015poor, waide2017demystifying, stark2018before, samuel2020machine, pineau2021improving, gundersen2022machine}.

The lack of detailed methodological reporting in DL studies has been a point of concern across multiple domains. Numerous researchers have called attention to the need for better documentation practices, emphasizing that insufficient details about model architecture, training procedures, and data preprocessing steps can lead to challenges in replicating results. For example, \cite{gundersen2018state} discusses how the reproducibility crisis has impacted other scientific disciplines and is now a growing concern in DL research due to these gaps in methodological transparency. Similarly, \cite{pineau2021improving} advocates for standardising reporting practices in machine learning papers to ensure that experiments can be independently reproduced, thereby fostering greater confidence in published findings.

In the biodiversity domain, DL has seen rapid adoption, driven by its potential to handle large-scale and complex ecological data. Applications of DL in biodiversity include species identification, habitat classification, and population monitoring, as evidenced by works such as \cite{christin2019applications} and \cite{norouzzadeh2017automatically}. 
However, inadequate documentation of DL methodologies is particularly problematic in biodiversity due to the field’s interdisciplinary nature and the complexity of ecological data. Biodiversity DL studies require transparency in methods to enable stakeholders, including conservationists and ecologists, to replicate and build upon findings effectively. The need for comprehensive and transparent methodological documentation in DL research is well-established, as studies without such documentation are difficult to reproduce or expand upon. In ecological and biodiversity studies, this problem is particularly acute, as the integration of DL methods is relatively new and is still evolving \citep{feng2019checklist}. \cite{whitlock2011data} and \cite{michener1997nongeospatial} previously raised the importance of archiving primary data with clear metadata to enhance reproducibility. More recently, efforts in ecological informatics have focused on creating reproducible workflows for studies involving complex DL techniques. By automating the extraction of categorical and structured responses from biodiversity DL publications, our pipeline addresses these reproducibility challenges, aiming to make DL methodologies in biodiversity research more accessible and consistent.
Addressing this need for methodological clarity is an important part of our work, which focuses on extracting DL methodologies from biodiversity publications. Our previous work has also emphasized these challenges \citep{ahmed2023reproducible,kommineni2024automatingtdwg}, especially in the context of semi-automated construction of the Knowledge Graphs (KGs) to improve data accessibility \citep{kommineni2024human}.

The emergence of Large Language Models has introduced new possibilities for automatically extracting and synthesizing information from text \citep{zhu2023large}, which can be particularly useful for addressing the gaps in methodological reporting. LLMs, such as GPT-3 \citep{brown2020language} and its successors \citep{achiam2023gpt,touvron2023llamaopenefficientfoundation,gemmateam2024gemma2improvingopen}, have demonstrated remarkable abilities in natural language understanding and generation, enabling tasks like summarization, question-answering, and information retrieval from vast textual datasets. Recent studies, including those by \citep{lewis2020retrieval} on Retrieval-Augmented Generation (RAG), have explored how combining LLMs with retrieval mechanisms can enhance the extraction of relevant information from large corpora, offering a promising solution for improving the accessibility of methodological details in scientific literature. In this study, we build on these developments by employing a multi-LLM and RAG-based pipeline to retrieve and categorize DL-related methodological details from scientific articles systematically.

While the application of LLMs for methodological extraction remains underexplored, several tools and approaches have been developed for automating information extraction \citep{beltagy2019scibert, lozano2023clinfo, dunn2022structured, dagdelen2024structured}. Tools like SciBERT \citep{beltagy2019scibert} and other domain-specific BERT models have been used to extract structured information from unstructured text, yet their application has primarily been focused on citation analysis, abstract summarization, or specific biomedical applications. \cite{bhaskar2024reproscreener} introduced "ReproScreener," a tool for evaluating computational reproducibility in machine learning pipelines, which uses LLMs to assess methodological consistency. Similarly, \cite{Gougherty2024TestingTR} tested an LLM-based approach for extracting ecological information, demonstrating the potential of LLMs to improve metadata reporting and transparency. These studies underscore the need for versatile, automated methodologies capable of handling DL pipeline documentation across various fields. 
The use of LLMs for extracting detailed DL methodologies across a broad spectrum of scientific domains remains underexplored. Our work aims to fill this gap by utilizing multiple LLMs in conjunction with a Retrieval-Augmented Generation (RAG) approach to extract and consolidate DL methodological details from biodiversity literature, offering a framework adaptable to other domains.

Recent studies have highlighted the environmental impact of computational processes, particularly in DL research. Training LLMs and executing complex pipelines can consume substantial energy, contributing to carbon emissions. \cite{lannelongue2021ten,lannelongue2021green} called for increased awareness of the ecological impact of computational research and proposed the routine assessment of environmental footprints as part of research best practices. 
In biodiversity research, where sustainability is a core value, these considerations are especially relevant. Our study contributes to this body of work by quantifying the environmental footprint of our DL-powered information retrieval pipeline using metrics such as kWh consumption and carbon emissions. This assessment is intended to encourage sustainable practices in computational research and aligns with recent recommendations to integrate environmental accountability into scientific workflows.


\section*{Methods}
In this section, we provide detailed information about the pipeline employed to extract and analyse the information from the selected biodiversity-related publications. 
\begin{figure}
\centering
\includegraphics[page=1, width=\textwidth]{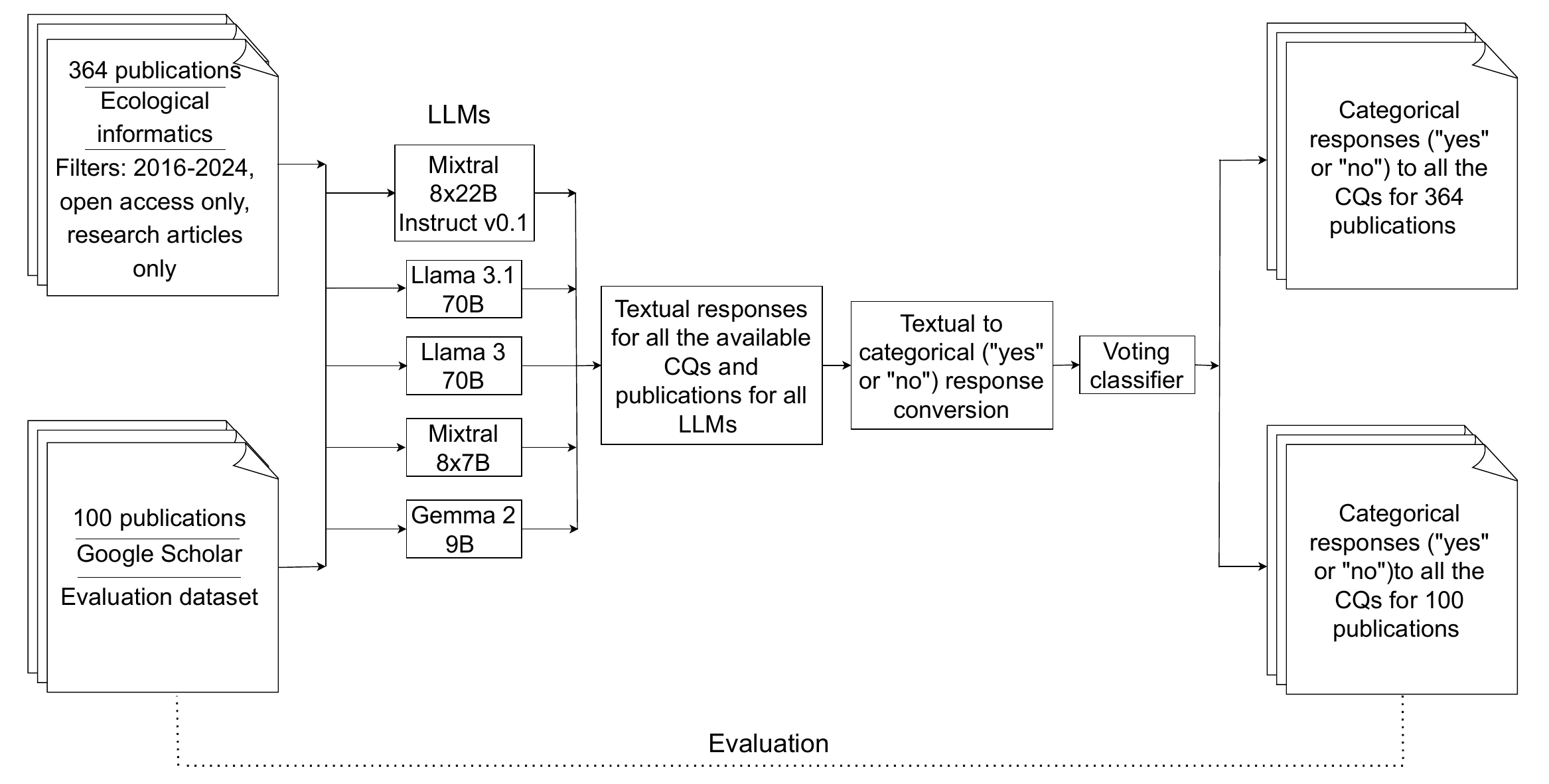}
\caption{Workflow of the pipeline. The solid arrows represent the main process flow, while the dotted line indicates the evaluation phase for the categorical responses from 100 publications retrieved from our previous research \citep{Peerj_repro_waq}.
}
\label{fig:methods}
\end{figure}
\subsection*{Dataset}
Our work is based on two datasets. The first one originates from our previous research \citep{Peerj_repro_waq}, while the second is sourced from the \textit{Ecological Informatics} Journal. Each dataset was indexed using different methodologies, contributing to a diverse representation of information. This variation arises from the range of journals included in the first dataset and the specific selection criteria applied in the second.
\subsubsection*{Dataset from Prior Research} \label{sec:previous_data}
In our previous study \citep{ahmed2024evaluating}, we used a modified version of the keywords from previous research \citep{abdelmageed2022biodivnere} to query Google Scholar and indexed over 8000 results. From this, the authors narrowed down the selection to 100 publications, excluding conference abstracts, theses, books, summaries, and preprints. Later, the first and second authors of that work manually extracted deep-learning information on ten variables (Dataset, Source Code, Open source frameworks or environment, Model architecture, Software and Hardware Specification, Methods, Hyperparameters, Randomness, Averaging result and Evaluation metrics) from the biodiversity publications, recording each as a categorical value: “yes” if the information was present and “no” if it was absent. In the current study, these 100 publications serve as an evaluation dataset, supporting the comparison and validation of our findings.
\subsubsection*{Dataset from \textit{Ecological Informatics} journal}
To index deep-learning-related publications from the \textit{Ecological Informatics} journal, we first identified relevant keywords and used them to guide the indexing of publications.\newline
\textbf{Keywords selection:} Related keywords are crucial for automatically indexing deep learning-related publications from a journal. To identify these relevant deep-learning keywords, we downloaded AI-related session abstracts from the Biodiversity Information Standards (TDWG) conferences\footnote{\url{https://www.tdwg.org/}} held in 2018 \citep{symai2018tdwg}, 2019 \citep{symai2019tdwg}, and 2021–2023 \citep{symai2021tdwg, symai2022tdwg, symai2023tdwg} (no AI session was available for 2020). We then used an open-source large language model (Mixtral 8x22b Instruct-v0.1) to extract all deep-learning-related keywords from each abstract. The query in the prompt template for extracting deep learning keywords from the given context is ``your task is to extract the deep learning related keywords from the provided context for the literature survey".
\begin{mdframed}[backgroundcolor=blue!5!white, linecolor=blue!75!black, 
topline=true, frametitle={\centering \textcolor{white}{Prompt for deep-learning-related keyword extraction}}, 
frametitlebackgroundcolor=blue!80!black, frametitlefont=\bfseries]
    ```

\%INSTRUCTIONS:

Use the provided pieces of context to answer the query. If you don't know the answer, just say that you don't know, don't try to make up an answer.

\%Query

Query: \{query\}

Context: \{context\}

Provide your answer as follows:

Answer:::

Deep learning related words: (Deep learning related words in comma separated list)

Answer:::

'''

\end{mdframed}

The LLM extracted a total of 248 keywords from 44 abstracts, averaging approximately 5.6 keywords per abstract. Since each abstract was treated individually during keyword extraction, the LLM indexed the same keywords multiple times, leading to redundancy and non-qualitative keywords. To improve keyword quality, we prompted the same LLM again with the full list, instructing it to eliminate redundancies and non-deep-learning-related terms. This refinement reduced the list from 248 to 123 keywords. Finally, a domain expert further curated this list down to 25 keywords (Figure \ref{fig:keywords})  by removing abbreviations and redundant terms, ensuring accurate indexing from the journal. \newline
\begin{figure}[h!]
    \centering
\labeledfbox{}{acoustic model, artificial intelligence, species identification, cnn model, convolutional neural network, convolutional vision transformer, deep learning, deep model, generative ai, handwritten text recognition, instance segmentation, large language model, machine learning, metric learning, natural language processing, neural network, object detection, object segmentation, optical character recognition, self-supervised learning, supervised learning, transfer learning, transformer, unsupervised learning, vision transformer} 
 \caption{LLM-Human optimized 25 DL-related keywords from 44 AI-related session abstracts at the Biodiversity Information Standards (TDWG) conferences.}
    \label{fig:keywords}
\end{figure}

\textbf{Publication citation data extraction:} Using the 25 refined keywords identified from TDWG abstracts with the assistance of both the LLM and domain experts, we queried the \textit{Ecological Informatics} journal.
The query applied the following filters: publication years from 2016 to August 1, 2024, article type as research articles, and open-access availability. Due to the platform’s limit of 8 boolean connectors per search, the keywords were divided into five sets, each connected with the boolean operator OR (e.g., "Keyword 1" OR "Keyword 2" OR "Keyword 3" OR "Keyword 4" OR "Keyword 5"). Citation data from each search was manually exported in BibTeX format. In total, 991 citation records were indexed, and after removing duplicates based on DOIs, 364 unique publications were identified.

\begin{figure}
\centering
\includegraphics[width=0.8\textwidth]{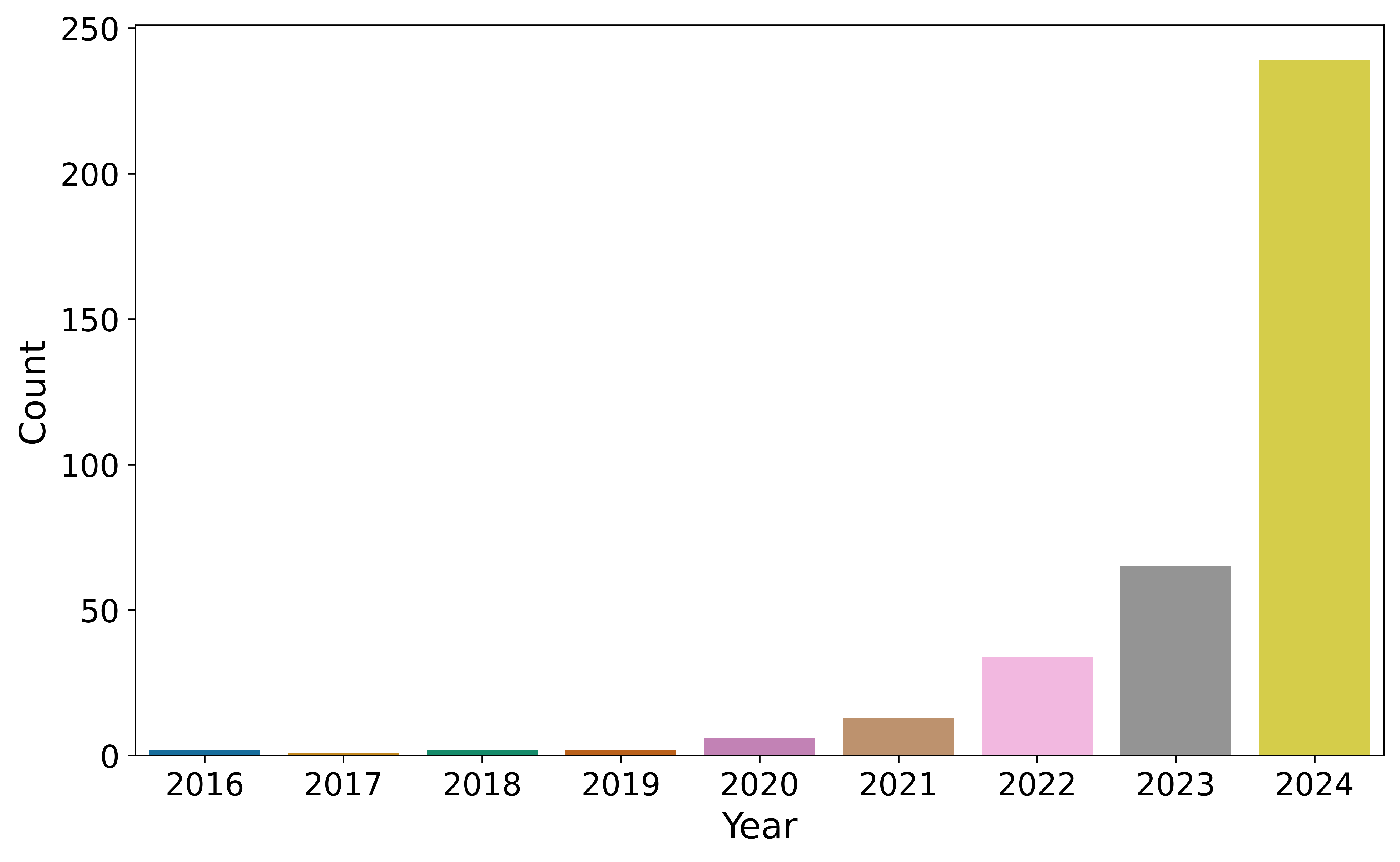}
\caption{Number of publications selected from \textit{Ecological Informatics} Journal (364 publications)}
\label{fig:year_barplot}
\end{figure}

The bar plot (Figure \ref{fig:year_barplot}) illustrates the annual distribution of these 364 publications from \textit{Ecological Informatics}. The trend shows a consistent increase in publication frequency up to 2023, with 65 data points recorded for that year. In 2024, there is a significant rise to 239 data points, representing a fourfold increase compared to 2023.  \newline
\textbf{Full-text publication download}:  Using the DOIs of the 364 unique publications, we retrieved the full-text PDFs through the Elsevier API. These PDFs were subsequently used as input for the selected LLMs.
\subsection*{Competency Questions (CQs)}
We employed competency questions (CQs) to retrieve specific deep-learning methodological information from selected biodiversity publications. Competency questions are natural language questions that users seek answers to and are essential for defining an ontology’s scope, purpose, and requirements \citep{gruninger1995role}. In our previous work \citep{kommineni2024human}, two domain experts formulated 28 CQs to cover every aspect of the DL pipeline for retrieving information from the provided context. For this study, we applied the same set of 28 CQs with multiple LLMs to extract relevant deep-learning information from a total of 464 biodiversity-related publications (364 from Ecological Informatics and 100 from previous research). 
\subsection*{Information retrieval}
Recently, the RAG approach has rapidly been used for information retrieval from both structured and unstructured data. This method leverages large language model (LLM) text generation to extract information from authoritative sources, such as biodiversity publications in our case. In this work, we employed five LLMs from two providers, namely hugging face Mixtral 8x22B Instruct v0.1\footnote{\url{https://huggingface.co/mistralai/Mixtral-8x22B-Instruct-v0.1}} and Groq's\footnote{\url{https://console.groq.com/docs/models}} Llama 3.1 70B, Llama 3 70B, Mixtral 8x7B and Gemma 2 9B with temperature set to 0 for all models. The Mixtral 8x22B Instruct v0.1 model was run on a custom GPU, while the Groq models were accessed through their API, where a custom GPU is not required. 

Information retrieval using LLMs and RAG was also a component of our previous work pipeline \citep{kommineni2024human}, where we aimed to build a semi-automated construction of the Knowledge Graph (KG) pipeline (we refer to the definition of KG from \citep{hogan2021knowledge}). This approach allowed us to extract, organize, and link information from unstructured text into structured, queryable data within the KG framework. By semi-automating the construction of KGs, we streamlined the process of mapping complex domain knowledge, which is crucial for advancing research in areas that require high levels of detail, such as biodiversity and deep learning methodologies. In this work, we build on our previous information retrieval component (then CQ Answering) by limiting the retrieval tokens to 1200, chunk size to 1000 and overlap to 50 chunks. Additionally, we specified that the responses should be concise and limited to fewer than 400 words to enhance the clarity and focus of the responses. 
For each selected LLM, the CQs and biodiversity-related publications were provided as input, and the RAG-assisted LLM pipeline generated answers to all CQ-publication combinations in textual sentence format as output.
\subsection*{Preprocessing LLM outputs}
After the information retrieval process, we obtained answers to the CQ for each combination of LLM, CQ, and publication. Some of these responses contained unnecessary structured information. To streamline the outputs, we preprocessed the responses using a Python script, removing strings like "Helpful Answer::" and "Answer::" to eliminate unnecessary content. We indexed only the information following these strings for the Mixtral 8x22B Instruct v0.1 model, as that portion contained details relevant to the queries and selected context.

Next, we converted all preprocessed LLM textual responses into categorical "yes" or "no" answers. To achieve this, we prompted the same LLMs that generated the CQ answers to provide categorical responses for each question-answer pair. To guide this process, a couple of in-context examples are provided in the prompt  as references for the LLM. This conversion from textual to categorical responses will later facilitate the evaluation of our pipeline.
\newpage
\begin{mdframed}[backgroundcolor=blue!5!white, linecolor=blue!75!black, 
topline=true, frametitle={\textcolor{white}{Prompt for converting textual to categorical responses}}, 
frametitlebackgroundcolor=blue!80!black, frametitlefont=\bfseries]

  ```
  
\%INSTRUCTIONS:

Your task is to determine if the provided answer contains specific information that could be derived from a research paper, based on the given question and answer. Provide a binary response ("yes" or "no"). Use the examples below as a reference for your decision.

"Yes": Choose "yes" if the answer includes specific details or direct information that seems likely to come from a research paper.

"No": Choose "no" if the answer is general or does not provide specific information that appears to be sourced from a research paper.

\%EXAMPLES:

Example 1:

Question: What methods are utilized for collecting raw data in the deep learning pipeline (e.g., surveys, sensors, public datasets)?

Answer: Unfortunately, there is no information provided about where the code repository of the deep learning pipeline is available. It could be hosted on platforms such as GitHub, GitLab, or BitBucket, but without explicit mention in the provided context, I cannot provide a definitive answer.

Response: No

Example 2:

Question: What data formats are used in the deep learning pipeline (e.g., image, audio, video, CSV)?

Answer: The study uses audio data from bird calls, specifically spectrograms derived from the audio files. These spectrograms serve as the input for the Convolutional Neural Network (CNN) model employed in the research. Therefore, the primary data format utilized in this deep learning pipeline is audio data, processed into spectrograms for further analysis.

Response: Yes

Example 3:

Question: What are the data annotation techniques used in the deep learning pipeline (e.g., bounding box annotation, instance segmentation)?

Answer: Unfortunately, there isn't any information about specific data annotation techniques used in the deep learning pipeline within the provided context. However, it does mention that Convolutional Neural Networks (CNNs) were employed for image classification tasks and pattern recognition. This implies that some form of image labeling or categorization was likely involved during the model training process. Nevertheless, without explicit details regarding the use of bounding box annotations, instance segmentations, or other similar techniques, I cannot provide a definitive response based solely on the given context.

Response: No

\%QUERY

Here are the "Question" and "Answer"

Question: \{Question\}

Answer: \{Answer\}

\%ANSWER FORMAT

Answer:::

Response: (Yes or No)

Answer:::

"""
\label{tclorbox:Prompt_textual_categorical}
\end{mdframed}

\subsection*{Voting classifier}
Since we leveraged multiple LLMs to retrieve the DL-related information and processed that information to categorical values, it became feasible to build a voting classifier. We employed a hard voting methodology, where each of the five instances (derived from five LLMs) produced possible outcomes of "yes" or "no" for each combination of CQ and publication. The voting classifier made decisions based on the majority of votes, which enhances the overall quality of the results.

\subsection*{Evaluation}
All key outputs generated by the LLMs, including the CQ answers and the conversion of textual responses to categorical values ("yes" or "no"), were manually evaluated. For assessing the CQ answers, we relied on our previous work \citep{kommineni2024human}, in which we manually evaluated 30 publications from the evaluation dataset. To evaluate the categorical responses produced by the LLMs, we randomly selected 30 publications, used those for each LLM, and manually annotated the ground truth data by assessing the question-answer pairs generated by the RAG-assisted LLM pipeline. We then compared the inter-annotator agreement between the LLM-generated and manually annotated answers using Cohen’s kappa score\footnote{\url{https://scikit-learn.org/stable/modules/generated/sklearn.metrics.cohen_kappa_score.html}}. This annotation process was conducted by the first and last authors of this paper.

\subsection*{Additional analysis}
\textbf{Publication filtering:} Our pipeline was driven by the DL-related keywords, which means that our dataset may include publications that mention these keywords without actually detailing a DL pipeline. To investigate this assumption as an addition to our current pipeline, we filtered the publications by using a RAG-assisted LLM pipeline (Llama 3.1 70B) to identify if any publications that contained only DL-related keywords, rather than discussing a DL pipeline. To evaluate the LLM's judgement, we compared its findings with 100 articles from our previous work \citep{Peerj_repro_waq}, where all the publications were focused on DL methods. Furthermore, we also compared the outputs of all the publications with those of filtered publications. \newline
\textbf{Time logs:} Computational tasks inherently rely on physical resources, and there is a growing awareness of the substantial environmental footprint associated with both the production and use of these resources \citep{lannelongue2021ten,samuel2024computational}. In the context of our work, which leverages information retrieval workflows involving DL methodologies in biodiversity research, one of our aims is to evaluate and quantify the environmental impact of these computational processes.
In this pipeline, we recorded the time taken to process all the requests for each document. We preprocessed the time logs by considering the last instance while removing the duplicates based on the unique identifiers of the log file. These time records are essential for calculating the environmental footprint \citep{lannelongue2021ten,lannelongue2021green} of the pipeline. By assessing the energy and resource consumption of our DL-driven information retrieval pipeline, we hope to contribute to more sustainable practices in biodiversity informatics and computational research more broadly. \newline
\textbf{Semantic similarity between five LLM outputs:} As mentioned before, we have five answers for each combination of CQ and publication—one from each LLM—formatted in both textual and categorical forms. We used these five textual answers to compute the cosine similarity\footnote{\url{https://scikit-learn.org/stable/modules/generated/sklearn.metrics.pairwise.cosine_similarity.html matrix}} matrix. With this matrix, average cosine similarities for all the responses between all the LLM combinations were calculated. Additionally, we assessed the inter-annotator agreement among the categorical responses using Cohen’s kappa score for all possible combinations. \newline
\textbf{Environmental footprint:}  Although our pipeline recorded processing times for each publication and each combination of CQ and publication, we only utilized the logged times for each publication for two key components of the pipeline (Table \ref{tab:Times}): 1. RAG answers and 2. Conversion of RAG textual responses to categorical responses. To estimate the environmental footprint, we used the website\footnote{\url{http://calculator.green-algorithms.org/}} \citep{lannelongue2021green}, which requires input on hardware configuration, total runtime, and location to estimate the environmental footprint of our computations. Our calculation only accounts for the pipeline components mentioned above and the hardware components from our side, excluding the hardware components from Groq. Our pipeline consumed 177.55 kWh of energy to generate the RAG textual responses, resulting in a carbon footprint of 60.14 kg CO2e, which is equivalent to the carbon offset of 64.65 tree months. For converting textual to categorical responses, the pipeline consumed 50.63 kWh of energy, corresponding to a carbon footprint of 17.15 kg CO2e and 18.7 tree months. For the environmental footprint estimates, we selected Germany as the location and assumed that we used the total number of cores in the Intel Xeon Platinum 9242 processor (which is 48 cores). 
\begin{table}[ht]
\centering
\begin{tabular}{p{2.25cm}p{3.75cm}p{3cm}p{3cm}}
\hline
\textbf{LLM name} &\textbf{Hardware} & \textbf{RAG textual responses} & \textbf{Conversion of textual to categorical responses}  \\ 
\hline
 Mixtral 8x22B Instruct v0.1& NVIDIA H100 (94 GB) & 71hr 3min& 6hr 34min\\ 
 & NVIDIA A100 (2x80 GB)& 69hr 10min & --\\ 
 Mixtral 8x7B &Intel Xeon Platinum 9242 &63hr 32min & 40hr 39min\\ 
Llama 3.1 70B & Intel Xeon Platinum 9242&5hr 52min &9hr 31min\\ 
Llama 3 70B &Intel Xeon Platinum 9242 &38hr 36min & 22hr 28min\\ 
Gemma 2 9B & Intel Xeon Platinum 9242&16hr 2min &8hr 49min \\ 
\hline
\bottomrule
\end{tabular}
\caption{Processing time for two key components of the pipeline}
\label{tab:Times}
\end{table}

\section*{Results}
This section presents the results from each part of the pipeline. We queried 28 CQs \citep{kommineni2024human} across 464 publications for each LLM, resulting in a total of 12,992 textual answers. Overall, we obtained 64,960 textual responses from the five selected LLMs. These textual responses were then converted into categorical "yes" or "no" responses using the respective LLMs.

To evaluate the LLM's judgements in these conversions, we compared the categorical responses against human-annotated ground truth data from 30 randomly selected publications. We used those randomly selected 30 publications for each LLM, leading to 840 comparisons per LLM (30 publications × 28 CQs).
This resulted in 4,200 comparisons for five LLMs, with 3,566 agreements between the LLM responses and the human-annotated ground truth responses, achieving a maximum agreement of 752 out of 840 for the Llama 3 70B model (Table \ref{tab:categorical_eval}).

The highest inter-annotator agreement between the LLM responses and human annotations was 0.7708, achieved with the Llama 3 70B model. This score reflects a strong level of agreement, as Scikit-learn's Cohen's Kappa score ranges from -1 (indicating no agreement) to +1 (indicating complete agreement) (Table \ref{tab:categorical_eval}).
\begin{table}[ht]
\centering
\begin{tabular}{|p{4cm}|p{4cm}|p{4cm}|}
\hline
\textbf{LLM name} & \textbf{Agreements between LLM and human response} & \textbf{Cohen’s kappa score (IAA)}  \\ \hline
 Mixtral 8x22B Instruct v0.1 & 667/840 & 0.5711\\ \hline
 Mixtral 8x7B &666/840 & 0.5583\\ \hline
Llama 3.1 70B &735/840 &0.7221\\ \hline
Llama 3 70B &752/840 & 0.7708\\ \hline
Gemma 2 9B & 746/840 &0.7128 \\ \hline
\end{tabular}
\caption{Evaluation of LLM responses when converting textual answers to categorical responses (``yes" or `no"). IAA = Inter-Annotator Agreement}
\label{tab:categorical_eval}
\end{table}  
As mentioned in the dataset subsection, we used a dataset from our previous work \citep{Peerj_repro_waq}, consisting of 100 publications, to evaluate our pipeline. We compared the manually annotated responses from that study \citep{Peerj_repro_waq} with the results generated by the voting classifier. Six DL reproducibility variables are both common to this work and the prior study, allowing us to analyze six CQs across 100 publications, which resulted in a total of 600 comparisons. 

There are 417 agreements between the human annotators from the previous work \citep{Peerj_repro_waq} and the voting classifier results. Table \ref{tab:mapping_cqs} shows the number of agreements between the human annotators and the voting classifier for each reproducibility variable. The DL variable \textit{Model architecture} has the highest agreement, with 89 agreements, while \textit{Open source framework} has the lowest, with 53 agreements. Table \ref{tab:mapping_cqs} also shows the mapping of CQs from this pipeline to the reproducibility variables of the previous work \citep{ahmed2024evaluating}.

\begin{table}[ht]
\centering
\begin{tabular}{|p{6.5cm}|p{3cm}|p{3cm}|}
\hline
\textbf{CQ} & \textbf{Deep Learning variable from \cite{ahmed2024evaluating}} &\textbf{Agreements between human response and voting classifier} \\ \hline
 What are the datasets used in the deep learning pipeline
(e.g., MNIST, CIFAR, ImageNet)? & Dataset &63/100 \\ \hline
What is the code repository link of the deep learning
pipeline (e.g., Link to GitHub, GitLab, BitBucket)? & Source code&74/100\\ \hline
Which frameworks are used to build the deep learning
model (e.g., TensorFlow, PyTorch)?&Open source framework&53/100\\ \hline
What type of deep learning model is used in the pipeline
(e.g., CNN, RNN, Transformer)?&Model architecture&89/100 \\ \hline
What are the hyperparameters used in the deep learning
model (e.g., learning rate, optimizer)?&Hyperparameters&63/100 \\ \hline
What metrics are used to evaluate the performance of the
deep learning model (e.g., accuracy, precision, recall)?&Metrics availability&75/100 \\ \hline
\end{tabular}
\caption{Mapping of current CQs to the deep learning variables in the previous work \citep{ahmed2024evaluating} and the number of agreements between the human annotators from \citep{ahmed2024evaluating} and the voting classifier for each reproducibility variable.}
\label{tab:mapping_cqs}
\end{table}
This serves as a proof of concept for validating the voting classifier for the remaining 364 publications (Table \ref{tab:mapping_cqs}). In this context, we calculated the voting classifier decisions for all 464 publications. After filtering out those publications that do not include a DL pipeline in their research, only 257 publications remained from the initial analysis (Table \ref{tab:voting_classifier_results}).  

Table \ref{tab:voting_classifier_results} shows that CQ 25 (\textit{purpose of the deep learning model}) is the most frequently mentioned, appearing in 345 publications. In contrast, CQ 27 (\textit{process to deploy the trained deep learning model}) is the least frequently mentioned. Following the filtering process, CQ 25 with 247 mentions, and CQ 27 with six mentions retain their positions as the most and least mentioned variables, respectively, among the 257 publications. With the current pipeline, 3,524 queries were answered out of a total of 12,992 total queries. After filtering the publications, 2,574 queries were answered out of 7,196 total queries.

\renewcommand{\arraystretch}{1.2} 
{\footnotesize   

\begin{longtable}{p{1cm} p{6.5cm} p{2cm} p{2.5cm}}
\caption{Number of publications providing information on specific CQs based on the voting classifier, before and after filtering out publications that do not include DL in the study.}
\label{tab:voting_classifier_results} \\
\toprule
\textbf{CQ Nr.} & \textbf{CQ} & \textbf{Number of publications that provide CQ info} & \textbf{Number of publications that provide CQ info after filtering the publications that do not contain DL in the study} \\
\midrule
\endfirsthead

\multicolumn{4}{c}{{\tablename\ \thetable{} -- Continued from previous page}} \\
\toprule
\textbf{CQ Nr.} & \textbf{CQ} & \textbf{Number of publications that provide CQ info} & \textbf{Number of publications that provide CQ info after filtering the publications that do not contain DL in the study} \\
\midrule
\endhead

\bottomrule
\multicolumn{4}{r}{{Continued on next page}} \\
\endfoot

\bottomrule
\endlastfoot

1  & What methods are utilized for collecting raw data in the deep learning pipeline (e.g., surveys, sensors, public datasets)?  & 215/464 
& 109/257\\ 
2  & What data formats are used in the deep learning pipeline (e.g., image, audio, video, CSV)? & 333/464 & 232/257 \\ 
3  & What are the data annotation techniques used in the deep learning pipeline (e.g., bounding box annotation, instance segmentation)?
 & 61/464 & 55/257 \\ 
4  & What are the data augmentation techniques applied in the deep learning pipeline (e.g., Flipping, Rotating, Scaling)? & 76/464 & 69/257 \\ 
5  & What are the datasets used in the deep learning pipeline (e.g., MNIST, CIFAR, ImageNet)?  & 152/464 & 134/257 \\ 
6  & What preprocessing steps are involved before training a deep learning model (e.g., normalization, scaling, cleaning)? & 145/464 & 92/257\\ 
7  & What are the criteria used to split the data for deep learning model training (e.g., train, test, validation)? & 141/464 & 102/257\\ 
8  & Where is the code repository of the deep learning pipeline available (e.g., GitHub, GitLab, BitBucket)? & 23/464 & 18/257\\  
9  & Where is the data repository of the deep learning pipeline available (e.g., Zenodo, Figshare, Dryad, GBIF)? & 27/464 & 16/257\\  
10 & What is the code repository link of the deep learning pipeline (e.g., Link to GitHub, GitLab, BitBucket)? & 20/464 & 17/257\\  
11 & What is the data repository link of the deep learning pipeline (e.g., Link to Zenodo, Figshare, Dryad, GBIF)? &  18/464 & 12/257\\  
12 & What type of deep learning model is used in the pipeline (e.g., CNN, RNN, Transformer)? & 275/464 & 235/257\\  
13 & What are the hyperparameters used in the deep learning model (e.g., learning rate, optimizer)? & 124/464 & 104/257\\  
14 & How are the hyperparameters of the model optimized (e.g., grid search, random search)? & 76/464 & 37/257\\  
15 & What optimization techniques are applied in the deep learning pipeline (e.g., SGD, Adam)? & 122/464 & 111/257\\  
16 & What criteria are used to determine when training is complete (e.g., validation loss plateau)? & 75/464 & 64/257\\  
17 & What are the regularization methods used to prevent overfitting in the deep learning pipeline (e.g., dropout, L2 regularization)? & 101/464 & 85/257\\  
18 & What is the strategy implemented to monitor the model performance during training? &  205/464& 129/257\\  
19 & Which frameworks are used to build the deep learning model (e.g., TensorFlow, PyTorch)? & 101/464 & 94/257\\  
20 & Which hardware resources are used for training the deep learning model (e.g., GPUs, TPUs)? & 101/464& 95/257 \\  
21 & What are the postprocessing steps involved after the model training (e.g., Saliency maps, Metrics calculation, Confusion matrix)? & 131/464 & 80/257 \\  
22 & What metrics are used to evaluate the performance of the deep learning model (e.g., accuracy, precision, recall)? & 340/464 & 225/257\\  
23 & What measures were taken to ensure the generalizability of the deep learning model (e.g., Diverse dataset, cross-validation, Stratified splitting)? &  174/464& 115/257\\  
24 & What strategies are employed to handle randomness in the deep learning pipeline (e.g., random seed value)?
 & 60/464 & 42/257\\  
25 & What is the purpose of the deep learning model (e.g., classification, segmentation, detection)?
 & 345/464 & 247/257\\  
26 & What techniques are used to address data bias during preprocessing of the deep learning pipeline (e.g., Stratified splitting, oversampling, undersampling, Diverse data collection)? &  59/464& 41/257\\  
27 & What process was followed to deploy the trained deep learning model (e.g., Model serialization, Platform selection)? & 7/464 & 6/257\\  
28 & Which platform was used to deploy the deep learning model (e.g., AWS, Azure, Google Cloud platform)? & 17/464 & 8/257\\  
\bottomrule
-- & Total for all queries & 3,524/12,992 & 2,574/7,196\\ 
\bottomrule
\end{longtable}
}

Additionally, we also computed the average cosine similarity scores for the RAG-assisted pipeline textual responses between different combinations of LLMs. This allows us to identify which LLM pairs provide similar outputs and assess whether different LLMs are generating comparable results. Table \ref{tab:Cosine_similarities} shows that the \textit{Llama 3.1 70B - Llama 3 70B} pair have the most similar answers, while \textit{Gemma 2 9B - Mixtral 8x22B Instruct v0.1} have the least similar answers before filtering. After filtering, the same LLM pairs perform in the same direction.   
\begin{table}[ht]
\centering
\begin{tabular}{|p{6.25cm}|p{2.5cm}|p{3.5cm}|}
\hline
\textbf{LLM pair} & \textbf{Cosine similarity score for all publications} & \textbf{Cosine similarity score after the removal of non-deep learning publications} \\ \hline
Gemma 2 9B - Llama 3.1 70B & 0.4619 &0.4857 \\ \hline
Gemma 2 9B - Llama 3 70B & 0.4773&0.5022\\ \hline
Gemma 2 9B - Mixtral 8x7B & 0.4201& 0.4327\\ \hline
Gemma 2 9B -  Mixtral 8x22B Instruct v0.1& 0.3989& 0.4128\\ \hline
Llama 3.1 70B - Llama 3 70B & 0.6854&0.6958\\ \hline
Llama 3.1 70B - Mixtral 8x7B & 0.5232& 0.5385\\ \hline
Llama 3.1 70B -  Mixtral 8x22B Instruct v0.1& 0.4759& 0.4959\\ \hline
Llama 3 70B - Mixtral 8x7B & 0.5374& 0.5505\\ \hline
Llama 3 70B -  Mixtral 8x22B Instruct v0.1& 0.4901&0.5064\\ \hline
Mixtral 8x7B -  Mixtral 8x22B Instruct v0.1& 0.4995& 0.5035\\ \hline
\end{tabular}
\caption{Average cosine similarity scores between all possible LLM pairs for the CQ textual responses}
\label{tab:Cosine_similarities}
\end{table}

Furthermore, the IAA scores were calculated for the categorical responses, which were generated from textual responses using LLMs for all the model combinations. The IAA score calculated using Scikit-learn Cohen’s Kappa score ranges from -1 (no agreement) to +1 (complete agreement). All calculated IAA scores range between 0.5321 and 0.7928, both inclusive, indicating moderate to strong agreement among all LLM pairs. Before the publication filtering, the Llama 3.1 70B - Llama 3 70B combination exhibits the maximum IAA score of 0.7924, while the Gemma 2 9B - Mixtral 8x7B combination has the minimum IAA score of 0.5321. After the filtering process, these same LLM pairs showed maximum and minimum IAA scores of 0.7928 and 0.5644 respectively (Table \ref{tab:IAA_category}). 
\begin{table}[ht]
\centering
\begin{tabular}{|p{6.25cm}|p{2.5cm}|p{3.5cm}|}
\hline
\textbf{LLM pair} & \textbf{Inter-Annotator Agreement score for all publications} & \textbf{Inter-Annotator Agreement score after the removal of non-deep learning publications} \\ \hline
Gemma 2 9B - Llama 3.1 70B & 0.6945 &0.7445 \\ \hline
Gemma 2 9B - Llama 3 70B & 0.6853&0.7312\\ \hline
Gemma 2 9B - Mixtral 8x7B & 0.5321& 0.5644\\ \hline
Gemma 2 9B -  Mixtral 8x22B Instruct v0.1& 0.6354& 0.6937\\ \hline
Llama 3.1 70B - Llama 3 70B & 0.7924&0.7928\\ \hline
Llama 3.1 70B - Mixtral 8x7B & 0.5533& 0.5784\\ \hline
Llama 3.1 70B -  Mixtral 8x22B Instruct v0.1& 0.6770& 0.7184\\ \hline
Llama 3 70B - Mixtral 8x7B & 0.5705& 0.5958\\ \hline
Llama 3 70B -  Mixtral 8x22B Instruct v0.1& 0.6901&0.7306\\ \hline
Mixtral 8x7B -  Mixtral 8x22B Instruct v0.1& 0.5581& 0.5992\\ \hline
\end{tabular}
\caption{Inter-Annotator Agreement scores between all possible LLM pairs for the CQ categorical responses.}
\label{tab:IAA_category}
\end{table}

\section*{Discussion}
Manually extracting DL-related information from scientific articles is both labour-intensive and time-consuming. Current approaches that rely on manual retrieval often vary significantly based on the annotator's perspective, which can differ from one annotator to another due to task interpretation and the annotators' domain knowledge \citep{ahmed2024evaluating}.
This variability can lead to inconsistencies and raises significant concerns regarding the reproducibility of manually annotated data.

To address these challenges, this work proposes an automated approach for retrieving information from scientific articles by employing five different LLMs. This strategy aims to improve both the accuracy and diversity of information extraction. By utilizing multiple LLMs, our pipeline is positioned to capture a broader range of variable-level information related to DL methodologies in scientific publications.

In this current pipeline, there are three critical components: 1. Identifying relevant research publications 2. Automatically extracting relevant information from publications for the desired queries, and 3. Converting the extracted textual responses into categorical responses. For the first component, we choose a method that extracts publications based on selected keywords. These keywords were derived from AI-related abstracts presented at the Biodiversity Information Standards (TDWG) conference, resulting in a total of 25 keywords. It is important to note that even if a publication mentions any of the keywords only once, without providing the actual DL methodology, it will still be included in the extraction process. As a result, our pipeline queries these publications, which may yield a higher number of negative responses, indicating that the context does not contain relevant information to answer the queries.

To mitigate this issue, we filtered the extracted publications again using the RAG-assisted pipeline. As a result, of this filtering, the number of publications decreased by 44.6\% , leaving us with 257 publications. This process was also evaluated using 100 publications from previous work \citep{Peerj_repro_waq}, all of which included DL methodologies in the study, and it achieved an accuracy of 93\%. 
Before filtering, our pipeline only provided positive responses to 27.12\% of the total queries (3,524 out of 12,992). After implementing the filtering step, the percentage of positive responses increased to 35.77\% (2,574 out of 7,196). This represents an improvement of 8.65\% in the positive response rate, which is a significant gain. However, after filtering, 64.23\% of the queries still did not yield available information in the publications. This gap can be attributed to the complexity of the queries (CQs), which cover all aspects of the DL pipeline, from data acquisition to model deployment.

In practice, not all studies utilize techniques like data augmentation; some prefer to use readily available datasets, thus bypassing the formal requirement for data annotation steps. Moreover, certain studies may not address model deployment at all. As a result, it is uncommon for publications to provide details on aspects such as deployment status, model randomness, generalizability, and other related factors. Consequently, the positive response rate for the queries tends to be relatively low.

To address the second component, we employed an RAG-assisted LLM pipeline to extract relevant information from the publications for all our queries (CQs). This component generated a total of 12,992 textual responses for each combination of queries (CQs) and publications across the different LLMs. The textual responses were initially preprocessed, and we calculated the average cosine similarity between the generated responses by different LLMs. The average cosine similarity score was high for the Llama 3.1 70B - Llama 3 70B model pair, indicating that these models generated similar outputs. On the other hand, the Gemma 2 9B - Mixtral 8x22B Instruct v0.1 model pair exhibited a lower average cosine similarity score, suggesting more significant variability in their response generation. Even after filtering the publications, the trend in the similarity scores remained consistent for these two model pairs, indicating that the response generation was not significantly affected by the exclusion of publications that did not utilize DL methods in their studies.

The third crucial component of our pipeline involves converting the extracted textual responses into categorical responses. This transformation simplifies the evaluation process, making it easier to compare the outputs generated by the LLM with human-extracted outputs from previous work \citep{Peerj_repro_waq}. Additionally, it facilitates the creation of an ensemble voting classifier. Two annotators reviewed the different question-answer pairs generated by the LLM and provided their assessments to ensure effective conversion from textual to categorical responses. The IAA scores between the human-annotated and LLM responses indicated that the highest levels of agreement were observed for the Llama 3 70B, Llama 3.1 70B, and Gemma 2 9B models in descending order, which generated straightforward answers that were easy for human annotators to evaluate. In contrast, the Mixtral 8x22B Instruct v0.1 and Mixtral 8x7B models exhibited the lowest IAA scores, reflecting only moderate agreement. The generated responses from these models were often ambiguous, combining actual answers with generalized or hallucinated content, which made it challenging for annotators to make precise judgments.

We also calculated the IAA scores for the categorical responses generated by different LLM pairs to evaluate the level of agreement among them. Overall, we observed a moderate to strong agreement between the various LLMs. However, following the publication filtering process, the IAA scores improved for all LLM pairs, indicating that the quality of the generated responses enhanced after the filtering.

The categorical responses have powered the ensemble approach of the voting classifier. We compared the categorical values from the voting classifier to the manually annotated values from our previous work for six deep-learning variables. This comparison revealed that the agreement between the LLM and human annotations is particularly low for the datasets, open-source frameworks, and hyperparameters. In the manual annotations, the authors from the previous work \citep{Peerj_repro_waq} also considered the accompanying code, which could explain the low agreement regarding open-source frameworks and hyperparameters. For datasets, the authors from the previous work \citep{Peerj_repro_waq} considered dataset availability only when persistent identifiers were provided in the respective studies. In contrast, the LLM also considers the dataset name itself, even when persistent identifiers are not mentioned.

Our approach incorporates a variety of models, each with distinct parameters, ensuring that the voting classifier considers diverse perspectives generated by different models for the same query. By ensembling the outputs of these varied models, the voting classifier enhances its robustness in making final decisions. This method not only enriches the decision-making process but also improves the classifier's overall reliability. 
\section*{Conclusions}
There is widespread concern about the lack of accessible methodological information in DL studies. We systematically evaluate whether that is the case for biodiversity research.  Our approach could be used to alleviate the problem in two ways: 1) by generating machine-accessible descriptions for a corpus of publications 2) by enabling authors and/or reviewers to verify methodological clarity in research papers.
Although our methodology has been demonstrated in the context of biodiversity studies, its applicability extends far beyond this field. It is a versatile approach that can be utilized across various scientific domains, particularly those where detailed, transparent, and reproducible methodological reporting is essential.


In this study, we used an automatic information retrieval method through an RAG-assisted LLM pipeline. Specifically, we employed five LLMs: Llama-3 70B, Llama-3.1 70B, Mixtral-8x22B-Instruct-v0.1, Mixtral 8x7B, and Gemma-2 9B to create an ensemble result, and then comparing the outputs with human responses from previous work \citep{ahmed2024evaluating}.
Our findings revealed that different LLMs generated varying outputs for the same query, indicating that information retrieval is not uniform across models. This underscores the necessity of considering multiple models to achieve more robust and accurate results. Additionally, precisely indexing publications that utilize DL methodologies significantly enhanced our results, and filtering out studies that did not employ these methods improved our findings. Furthermore, our results demonstrated that incorporating multiple modalities enriched the retrieval process, as evidenced by comparisons between the outputs of previous work \citep{ahmed2024evaluating} and our study's outputs.
 

In future research, we plan to develop a hybrid system that combines human expertise with LLM capabilities, where the LLMs will evaluate results using a metric to ensure the accuracy of generated outputs. In instances where the metric score is low, humans will manually assess those cases. We also aim to include different modalities (such as code and figures) in the pipeline to ensure more accurate information retrieval.
\section*{Acknowledgments}
Supported by the German Centre for Integrative Biodiversity Research (iDiv) Halle-Jena-Leipzig, which is funded by the German Research Foundation (DFG) under FZT 118 (ID 202548816) and also supported by the DFG project TRR 386 (ID 514664767). 
The authors gratefully acknowledge the computing time granted by the Resource Allocation Board and provided on the supercomputer Emmy/Grete at NHR@Göttingen as part of the NHR infrastructure. The calculations for this research were conducted with computing resources under the project nhr\_th\_starter\_22233.

\section*{Data availability statement}
The Data and Code used in this study are available in GitHub: \url{https://github.com/fusion-jena/information-retrieval-using-multiple-LLM-and-RAG}. 
\bibliography{main}

\begin{thebibliography}{}

\bibitem[Abdelmageed et~al., 2022]{abdelmageed2022biodivnere}
Abdelmageed, N., L{\"o}ffler, F., Feddoul, L., Algergawy, A., Samuel, S.,
  Gaikwad, J., Kazem, A., and K{\"o}nig-Ries, B. (2022).
\newblock {BiodivNERE}: Gold standard corpora for named entity recognition and
  relation extraction in the biodiversity domain.
\newblock {\em Biodiversity Data Journal}, 10.

\bibitem[Achiam et~al., 2023]{achiam2023gpt}
Achiam, J., Adler, S., Agarwal, S., Ahmad, L., Akkaya, I., Aleman, F.~L.,
  Almeida, D., Altenschmidt, J., Altman, S., Anadkat, S., et~al. (2023).
\newblock Gpt-4 technical report.
\newblock {\em arXiv preprint arXiv:2303.08774}.

\bibitem[Ahmed et~al., 2023]{ahmed2023reproducible}
Ahmed, W., Kommineni, V.~K., Koenig-ries, B., and Samuel, S. (2023).
\newblock How reproducible are the results gained with the help of deep
  learning methods in biodiversity research?
\newblock {\em Biodiversity Information Science and Standards}, 7.

\bibitem[Ahmed et~al., 2024a]{Peerj_repro_waq}
Ahmed, W., Kommineni, V.~K., K{\"o}nig-Ries, B., Gaikwad, J., Gadelha, L., and
  Samuel, S. (2024a).
\newblock Evaluating the method reproducibility of deep learning models in the
  biodiversity research.
\newblock {\em PeerJ Computer Science, Under review}.

\bibitem[Ahmed et~al., 2024b]{ahmed2024evaluating}
Ahmed, W., Kommineni, V.~K., K{\"{o}}nig{-}Ries, B., Gaikwad, J., Jr., L. M.
  R.~G., and Samuel, S. (2024b).
\newblock Evaluating the method reproducibility of deep learning models in the
  biodiversity domain.
\newblock {\em CoRR}, abs/2407.07550.

\bibitem[August et~al., 2020]{AUGUST2020100116}
August, T.~A., Pescott, O.~L., Joly, A., and Bonnet, P. (2020).
\newblock {AI} naturalists might hold the key to unlocking biodiversity data in
  social media imagery.
\newblock {\em Patterns}, 1(7):100116.

\bibitem[Beltagy et~al., 2019]{beltagy2019scibert}
Beltagy, I., Lo, K., and Cohan, A. (2019).
\newblock {S}ci{BERT}: A pretrained language model for scientific text.
\newblock In Inui, K., Jiang, J., Ng, V., and Wan, X., editors, {\em
  Proceedings of the 2019 Conference on Empirical Methods in Natural Language
  Processing and the 9th International Joint Conference on Natural Language
  Processing (EMNLP-IJCNLP)}, pages 3615--3620, Hong Kong, China. Association
  for Computational Linguistics.

\bibitem[Bhaskar and Stodden, 2024]{bhaskar2024reproscreener}
Bhaskar, A. and Stodden, V. (2024).
\newblock Reproscreener: Leveraging llms for assessing computational
  reproducibility of machine learning pipelines.
\newblock In {\em Proceedings of the 2nd ACM Conference on Reproducibility and
  Replicability}, ACM REP '24, page 101–109, New York, NY, USA. Association
  for Computing Machinery.

\bibitem[Brown et~al., 2020]{brown2020language}
Brown, T.~B., Mann, B., Ryder, N., Subbiah, M., Kaplan, J., Dhariwal, P.,
  Neelakantan, A., Shyam, P., Sastry, G., Askell, A., Agarwal, S.,
  Herbert{-}Voss, A., Krueger, G., Henighan, T., Child, R., Ramesh, A.,
  Ziegler, D.~M., Wu, J., Winter, C., Hesse, C., Chen, M., Sigler, E., Litwin,
  M., Gray, S., Chess, B., Clark, J., Berner, C., McCandlish, S., Radford, A.,
  Sutskever, I., and Amodei, D. (2020).
\newblock Language models are few-shot learners.
\newblock {\em CoRR}, abs/2005.14165.

\bibitem[Christin et~al., 2019]{christin2019applications}
Christin, S., Hervet, {\'E}., and Lecomte, N. (2019).
\newblock Applications for deep learning in ecology.
\newblock {\em Methods in Ecology and Evolution}, 10(10):1632--1644.

\bibitem[Dagdelen et~al., 2024]{dagdelen2024structured}
Dagdelen, J., Dunn, A., Lee, S., Walker, N., Rosen, A.~S., Ceder, G., Persson,
  K.~A., and Jain, A. (2024).
\newblock Structured information extraction from scientific text with large
  language models.
\newblock {\em Nature Communications}, 15(1):1418.

\bibitem[Dunn et~al., 2022]{dunn2022structured}
Dunn, A., Dagdelen, J., Walker, N., Lee, S., Rosen, A.~S., Ceder, G., Persson,
  K., and Jain, A. (2022).
\newblock Structured information extraction from complex scientific text with
  fine-tuned large language models.
\newblock {\em arXiv preprint arXiv:2212.05238}.

\bibitem[El-Amir and Hamdy, 2020]{el2020deep}
El-Amir, H. and Hamdy, M. (2020).
\newblock Deep learning pipeline.
\newblock {\em Apress: Berkeley, CA, USA}.

\bibitem[Feng et~al., 2019]{feng2019checklist}
Feng, X., Park, D.~S., Walker, C., Peterson, A.~T., Merow, C., and Pape{\c{s}},
  M. (2019).
\newblock A checklist for maximizing reproducibility of ecological niche
  models.
\newblock {\em Nature Ecology \& Evolution}, 3(10):1382--1395.

\bibitem[Frandsen et~al., 2019]{symai2019tdwg}
Frandsen, P., Dikow, R., Trizna, M., and White, A. (2019).
\newblock {SS86 - Machine learning: an emerging toolkit for biodiversity
  science using museum collections}.
\newblock {\em Biodiversity Information Science and Standards}.

\bibitem[Gougherty and Clipp, 2024]{Gougherty2024TestingTR}
Gougherty, A.~V. and Clipp, H.~L. (2024).
\newblock Testing the reliability of an ai-based large language model to
  extract ecological information from the scientific literature.
\newblock {\em npj Biodiversity}, 3.

\bibitem[GPAI, 2022]{gpai2022biodiversity}
GPAI (2022).
\newblock Biodiversity and artificial intelligence, opportunities and
  recommendations report.

\bibitem[Groom and Ellwood, 2021]{symai2021tdwg}
Groom, Q. and Ellwood, E. (2021).
\newblock {SYM01 - Applications of machine learning in biodiversity image
  analysis}.
\newblock {\em Biodiversity Information Science and Standards}.

\bibitem[Gr{\"u}ninger and Fox, 1995]{gruninger1995role}
Gr{\"u}ninger, M. and Fox, M.~S. (1995).
\newblock The role of competency questions in enterprise engineering.
\newblock In {\em Benchmarking—Theory and practice}, pages 22--31. Springer.

\bibitem[Gundersen and Kjensmo, 2018]{gundersen2018state}
Gundersen, O.~E. and Kjensmo, S. (2018).
\newblock State of the art: Reproducibility in artificial intelligence.
\newblock {\em Proceedings of the AAAI Conference on Artificial Intelligence},
  32(1).

\bibitem[Gundersen et~al., 2022]{gundersen2022machine}
Gundersen, O.~E., Shamsaliei, S., and Isdahl, R.~J. (2022).
\newblock Do machine learning platforms provide out-of-the-box reproducibility?
\newblock {\em Future Generation Computer Systems}, 126:34--47.

\bibitem[Haddaway and Verhoeven, 2015]{haddaway2015poor}
Haddaway, N.~R. and Verhoeven, J.~T. (2015).
\newblock Poor methodological detail precludes experimental repeatability and
  hampers synthesis in ecology.
\newblock {\em Ecology and Evolution}, 5(19):4451--4454.

\bibitem[Hogan et~al., 2021]{hogan2021knowledge}
Hogan, A., Blomqvist, E., Cochez, M., d’Amato, C., Melo, G.~D., Gutierrez,
  C., Kirrane, S., Gayo, J. E.~L., Navigli, R., Neumaier, S., et~al. (2021).
\newblock Knowledge graphs.
\newblock {\em ACM Computing Surveys (Csur)}, 54(4):1--37.

\bibitem[Johaadien et~al., 2023]{symai2023tdwg}
Johaadien, R., Lewers, K., and Torma, M. (2023).
\newblock {SYM05 AI Contributions to biodiversity data \& data standardisation:
  Opportunities and challenges}.
\newblock {\em Biodiversity Information Science and Standards}.

\bibitem[Kommineni et~al., 2024a]{kommineni2024automatingtdwg}
Kommineni, V.~K., Ahmed, W., Koenig-Ries, B., and Samuel, S. (2024a).
\newblock Automating information retrieval from biodiversity literature using
  large language models: A case study.
\newblock {\em Biodiversity Information Science and Standards}, 8:e136735.

\bibitem[Kommineni et~al., 2022]{symai2022tdwg}
Kommineni, V.~K., Groom, Q., and Panda, R. (2022).
\newblock {SYM12 - Information extraction from digital specimen images using
  Artificial Intelligence}.
\newblock {\em Biodiversity Information Science and Standards}.

\bibitem[Kommineni et~al., 2024b]{kommineni2024human}
Kommineni, V.~K., K{\"{o}}nig{-}Ries, B., and Samuel, S. (2024b).
\newblock From human experts to machines: An {LLM} supported approach to
  ontology and knowledge graph construction.
\newblock {\em CoRR}, abs/2403.08345.

\bibitem[Lannelongue et~al., 2021a]{lannelongue2021ten}
Lannelongue, L., Grealey, J., Bateman, A., and Inouye, M. (2021a).
\newblock Ten simple rules to make your computing more environmentally
  sustainable.
\newblock {\em PLOS Computational Biology}, 17(9):e1009324.

\bibitem[Lannelongue et~al., 2021b]{lannelongue2021green}
Lannelongue, L., Grealey, J., and Inouye, M. (2021b).
\newblock Green algorithms: Quantifying the carbon footprint of computation.
\newblock {\em Advanced Science}, 8(12):2100707.

\bibitem[Lewis et~al., 2020]{lewis2020retrieval}
Lewis, P. S.~H., Perez, E., Piktus, A., Petroni, F., Karpukhin, V., Goyal, N.,
  K{\"{u}}ttler, H., Lewis, M., Yih, W., Rockt{\"{a}}schel, T., Riedel, S., and
  Kiela, D. (2020).
\newblock Retrieval-augmented generation for knowledge-intensive {NLP} tasks.
\newblock {\em CoRR}, abs/2005.11401.

\bibitem[Lozano et~al., 2023]{lozano2023clinfo}
Lozano, A., Fleming, S.~L., Chiang, C.-C., and Shah, N. (2023).
\newblock Clinfo. ai: An open-source retrieval-augmented large language model
  system for answering medical questions using scientific literature.
\newblock In {\em PACIFIC SYMPOSIUM ON BIOCOMPUTING 2024}, pages 8--23. World
  Scientific.

\bibitem[Michener et~al., 1997]{michener1997nongeospatial}
Michener, W.~K., Brunt, J.~W., Helly, J.~J., Kirchner, T.~B., and Stafford,
  S.~G. (1997).
\newblock Nongeospatial metadata for the ecological sciences.
\newblock {\em Ecological Applications}, 7(1):330--342.

\bibitem[Norouzzadeh et~al., 2017]{norouzzadeh2017automatically}
Norouzzadeh, M.~S., Nguyen, A.~T., Kosmala, M., Swanson, A., Palmer, M.~S.,
  Packer, C., and Clune, J. (2017).
\newblock Automatically identifying, counting, and describing wild animals in
  camera-trap images with deep learning.
\newblock {\em Proceedings of the National Academy of Sciences of the United
  States of America}, 115:E5716 -- E5725.

\bibitem[Pando et~al., 2018]{symai2018tdwg}
Pando, F., Mata, E., Carranza-Rojas, J., Lloret, L., and GOEAU, H. (2018).
\newblock {W14 Deep Learning for Biodiversity}.
\newblock {\em Biodiversity Information Science and Standards}.

\bibitem[Pineau et~al., 2021]{pineau2021improving}
Pineau, J., Vincent-Lamarre, P., Sinha, K., Larivi{\`e}re, V., Beygelzimer, A.,
  d'Alch{\'e} Buc, F., Fox, E., and Larochelle, H. (2021).
\newblock Improving reproducibility in machine learning research (a report from
  the neurips 2019 reproducibility program).
\newblock {\em The Journal of Machine Learning Research}, 22(1):7459--7478.

\bibitem[Samuel et~al., 2021]{samuel2020machine}
Samuel, S., L{\"{o}}ffler, F., and K{\"{o}}nig{-}Ries, B. (2021).
\newblock Machine learning pipelines: Provenance, reproducibility and {FAIR}
  data principles.
\newblock In Glavic, B., Braganholo, V., and Koop, D., editors, {\em Provenance
  and Annotation of Data and Processes - 8th and 9th International Provenance
  and Annotation Workshop, {IPAW} 2020 + {IPAW} 2021, Virtual Event, July
  19-22, 2021, Proceedings}, volume 12839 of {\em Lecture Notes in Computer
  Science}, pages 226--230. Springer.

\bibitem[Samuel and Mietchen, 2024]{samuel2024computational}
Samuel, S. and Mietchen, D. (2024).
\newblock Computational reproducibility of jupyter notebooks from biomedical
  publications.
\newblock {\em GigaScience}, 13:giad113.

\bibitem[Stark, 2018]{stark2018before}
Stark, P.~B. (2018).
\newblock Before reproducibility must come preproducibility.
\newblock {\em Nature}, 557(7706):613--614.

\bibitem[Team et~al., 2024]{gemmateam2024gemma2improvingopen}
Team, G., Riviere, M., Pathak, S., Sessa, P.~G., Hardin, C., Bhupatiraju, S.,
  Hussenot, L., Mesnard, T., Shahriari, B., Ramé, A., Ferret, J., Liu, P.,
  Tafti, P., Friesen, A., Casbon, M., Ramos, S., Kumar, R., Lan, C.~L., Jerome,
  S., Tsitsulin, A., Vieillard, N., Stanczyk, P., Girgin, S., Momchev, N.,
  Hoffman, M., Thakoor, S., Grill, J.-B., Neyshabur, B., Bachem, O., Walton,
  A., Severyn, A., Parrish, A., Ahmad, A., Hutchison, A., Abdagic, A., Carl,
  A., Shen, A., Brock, A., Coenen, A., Laforge, A., Paterson, A., Bastian, B.,
  Piot, B., Wu, B., Royal, B., Chen, C., Kumar, C., Perry, C., Welty, C.,
  Choquette-Choo, C.~A., Sinopalnikov, D., Weinberger, D., Vijaykumar, D.,
  Rogozińska, D., Herbison, D., Bandy, E., Wang, E., Noland, E., Moreira, E.,
  Senter, E., Eltyshev, E., Visin, F., Rasskin, G., Wei, G., Cameron, G.,
  Martins, G., Hashemi, H., Klimczak-Plucińska, H., Batra, H., Dhand, H.,
  Nardini, I., Mein, J., Zhou, J., Svensson, J., Stanway, J., Chan, J., Zhou,
  J.~P., Carrasqueira, J., Iljazi, J., Becker, J., Fernandez, J., van
  Amersfoort, J., Gordon, J., Lipschultz, J., Newlan, J., yeong Ji, J.,
  Mohamed, K., Badola, K., Black, K., Millican, K., McDonell, K., Nguyen, K.,
  Sodhia, K., Greene, K., Sjoesund, L.~L., Usui, L., Sifre, L., Heuermann, L.,
  Lago, L., McNealus, L., Soares, L.~B., Kilpatrick, L., Dixon, L., Martins,
  L., Reid, M., Singh, M., Iverson, M., Görner, M., Velloso, M., Wirth, M.,
  Davidow, M., Miller, M., Rahtz, M., Watson, M., Risdal, M., Kazemi, M.,
  Moynihan, M., Zhang, M., Kahng, M., Park, M., Rahman, M., Khatwani, M., Dao,
  N., Bardoliwalla, N., Devanathan, N., Dumai, N., Chauhan, N., Wahltinez, O.,
  Botarda, P., Barnes, P., Barham, P., Michel, P., Jin, P., Georgiev, P.,
  Culliton, P., Kuppala, P., Comanescu, R., Merhej, R., Jana, R., Rokni, R.~A.,
  Agarwal, R., Mullins, R., Saadat, S., Carthy, S.~M., Cogan, S., Perrin, S.,
  Arnold, S. M.~R., Krause, S., Dai, S., Garg, S., Sheth, S., Ronstrom, S.,
  Chan, S., Jordan, T., Yu, T., Eccles, T., Hennigan, T., Kocisky, T., Doshi,
  T., Jain, V., Yadav, V., Meshram, V., Dharmadhikari, V., Barkley, W., Wei,
  W., Ye, W., Han, W., Kwon, W., Xu, X., Shen, Z., Gong, Z., Wei, Z., Cotruta,
  V., Kirk, P., Rao, A., Giang, M., Peran, L., Warkentin, T., Collins, E.,
  Barral, J., Ghahramani, Z., Hadsell, R., Sculley, D., Banks, J., Dragan, A.,
  Petrov, S., Vinyals, O., Dean, J., Hassabis, D., Kavukcuoglu, K., Farabet,
  C., Buchatskaya, E., Borgeaud, S., Fiedel, N., Joulin, A., Kenealy, K.,
  Dadashi, R., and Andreev, A. (2024).
\newblock Gemma 2: Improving open language models at a practical size.

\bibitem[Touvron et~al., 2023]{touvron2023llamaopenefficientfoundation}
Touvron, H., Lavril, T., Izacard, G., Martinet, X., Lachaux, M.-A., Lacroix,
  T., Rozière, B., Goyal, N., Hambro, E., Azhar, F., Rodriguez, A., Joulin,
  A., Grave, E., and Lample, G. (2023).
\newblock Llama: Open and efficient foundation language models.

\bibitem[Waide et~al., 2017]{waide2017demystifying}
Waide, R.~B., Brunt, J.~W., and Servilla, M.~S. (2017).
\newblock {Demystifying the landscape of ecological data repositories in the
  United States}.
\newblock {\em BioScience}, 67(12):1044--1051.

\bibitem[Whitlock, 2011]{whitlock2011data}
Whitlock, M.~C. (2011).
\newblock Data archiving in ecology and evolution: best practices.
\newblock {\em Trends in ecology \& evolution}, 26(2):61--65.

\bibitem[Zhu et~al., 2023]{zhu2023large}
Zhu, Y., Yuan, H., Wang, S., Liu, J., Liu, W., Deng, C., Chen, H., Dou, Z., and
  Wen, J.-R. (2023).
\newblock Large language models for information retrieval: A survey.
\newblock {\em arXiv preprint arXiv:2308.07107}.

\end{thebibliography}

\end{document}